\documentclass[prb,twocolumn,showpacs,amsmath,amssymb,superscriptaddress]{revtex4}
\usepackage{epsfig}
\usepackage{graphicx}
\usepackage{amsmath,amssymb}

\newcommand{\beq}[1]{
\begin{equation}
\label{e#1} }

\newcommand{\eeq}{
\end{equation}
}

\begin{document}

\title{CuMn-V compounds: a transition from semimetal low-temperature to semiconductor high-temperature antiferromagnets }

\author{F.~M\'aca}
\author{J.~Ma\v{s}ek}
\affiliation{Institute of Physics ASCR, v.v.i., Na Slovance 2, 182 21 Praha 8, Czech Republic}
\author{O.~Stelmakhovych}
\author{X.~Mart\'{i}}
\author{K.~Uhl\'{\i}\v{r}ov\'a}
\affiliation{Faculty of Mathematics and Physics, Charles University in Prague, Ke Karlovu 3, 121 16 Prague 2, Czech Republic}
\author{P.~Beran}
\affiliation{Nuclear Physics Institute ASCR, v.v.i., 250 68 \v{R}e\v{z}, Czech Republic}
\author{H. Reichlov\'a}
\affiliation{Institute of Physics ASCR, v.v.i., Cukrovarnick\'a 10, 162 53 Praha 6, Czech Republic}
%\author{V.~Hol\'{y}}
%\affiliation{Faculty of Mathematics and Physics, Charles University in Prague, Ke Karlovu 3, 121 16 Prague 2, Czech Republic}
\author{P. Wadley}
\author{V. Nov\'ak}
\affiliation{Institute of Physics ASCR, v.v.i., Cukrovarnick\'a 10, 162 53 Praha 6, Czech Republic}
\author{T.~Jungwirth}
\affiliation{Institute of Physics ASCR, v.v.i., Cukrovarnick\'a 10, 162 53 Praha 6, Czech Republic}
\affiliation{School of Physics and Astronomy, University of Nottingham, Nottingham NG7 2RD, United Kingdom}

\begin{abstract}
We report on a theoretical and experimental study of CuMn-V antiferromagnets. Previous works showed low-temperature antiferomagnetism and semimetal electronic structure of the semi-Heusler CuMnSb. In this paper we present  theoretical predictions of high-temperature antiferromagnetism in the stable orthorhombic phases of CuMnAs and CuMnP. The electronic structure of CuMnAs is at the transition from a semimetal to a semiconductor and we predict that CuMnP is a semiconductor. We show that the transition to a semiconductor-like band structure upon introducing the lighter group-V elements is present in both the metastable semi-Heusler and the stable orthorhombic crystal structures. On the other hand, the orthorhombic phase is crucial for the high N\'eel temperature. Results of X-ray diffraction, magnetization, transport, and neutron diffraction measurements we performed on chemically synthesized CuMnAs are consistent with the theory predictions. 
 \end{abstract}

\date{\today}

\maketitle

\section{Introduction}
Recent observation of a large magnetoresistance in an antiferromagnet (AFM) based spin-valve opens the prospect for utilizing AFMs in spintronics.\cite{Shick:2010_a,Park:2010_a} Particularly appealing is the introduction of AFMs into semiconductor spintronics because of the lack of suitable high-temperature ferromagnetic (FM) semiconductors.  In Tab~\ref{tab1} we show a survey of the magnetic counterparts of the most common II-VI and III-V compound semiconductors, and of the related I-VI-III-VI and II-V-IV-V families, in which Mn (Eu) acts as a group-II atom and Fe (Gd)  as a group-III element. The table illustrates that AFM ordering occurs much more frequently than FM ordering. Yet, only a few of these AFM semiconductors have N\'eel temperatures $T_N$ above the room temperature. In MnTe, $T_N=323$~K is presumably still too low to allow for room-temperature applications of the material in spintronics. MnSiN$_2$ appears as an attractive candidate material which should also allow for the application of  common molecular beam epitaxy techniques for the synthesis of high quality films. The natural mineral CuFeS$_2$ is more challenging from the perspective of the epitaxial growth because of the vastly different vapor pressures of S and the noble and transition metals. Another limiting factor is that both MnSiN$_2$ and CuFeS$_2$ might be the only high-$T_N$ AFMs in their respective semiconductor compound families.

\begin{table}[h]
\begin{center}
\begin{tabular}{cccc|cccc}
\hline
II-VI & $T_c$~(K) & $T_N$~(K) &  & III-V & $T_c$~(K) & $T_N$~(K) &\\
%\hline
%\hline
MnO & &122&\cite{Nagaev:1975_a}&FeN& &100 &\cite{Suziki:1993_a}\\
MnS& &152&\cite{Chen:2009_a}&FeP& &115&\cite{Westerstrandh:1977_a}\\
MnSe& &173& \cite{Nagaev:1975_a}&FeAs& &77&\cite{Selte:1972_a}\\
MnTe& &{\bf 323}& \cite{Nagaev:1975_a}&FeSb& &100-220&\cite{Picone:1981_a}\\
%\hline
EuO&67& & \cite{Nagaev:1975_a}&GdN&72& &\cite{Nagaev:1975_a}\\
EuS&16& & \cite{Nagaev:1975_a}&GdP& &15&\cite{Kaldis:1975_a}\\
EuSe& &5& \cite{Nagaev:1975_a}&GdAs& &19&\cite{Li:1996_a}\\
EuTe& &10& \cite{Nagaev:1975_a}&GdSb& &27&\cite{Missell:1977_a}\\
\hline
%\hline
I-VI-III-VI& & & &II-V-IV-V & & &\\
CuFeO$_2$& &11& \cite{Mekata:1993_a}&MnSiN$_2$& &{\bf 490}&\cite{Esmaeilzadeh:2006_a}\\
CuFeS$_2$& &{\bf 825}& \cite{Nagaev:1975_a}& & & &\\
CuFeSe$_2$& &70& \cite{Lamazares:1992_a}& & & &\\
CuFeTe$_2$& &254& \cite{Rivas:1998_a}& & & &\\

\hline
\end{tabular}
\label{table1}
\end{center}
\caption{Comparison of FM Curie temperatures ($T_c$) and AFM N\'eel temperatures ($T_N$) of II-VI, I-VI-III-VI, III-V, and II-V-IV-V magnetic semiconductors.}
\label{tab1}
\end{table}

The search for other high temperature AFM semiconductors has recently resulted in a report of the semiconducting band structure  of  I(a)-Mn-V compounds and of the successful synthesis of single-crystal LiMnAs by molecular beam epitaxy.\cite{Jungwirth:2010_a}  In contrast to the other common semiconductor compound families, many of the I(a)-Mn-V semiconductors are room-temperature AFMs.\cite{Bronger:1986_a,Schucht:1999_a} While favorable from the perspective of their electronic band structure and magnetic characteristics, the utility of I(a)-Mn-V materials in devices may represent a challenge due to the high reactivity and diffusivity of the I(a) alkali metals elements. The aim of this paper is to investigate the noble-metal group-I(b) counterparts of these compounds. In particular, we focus on CuMnAs and CuMnP.

The anticipation of semiconducting band structure of the high-temperature AFM LiMnAs (and other I(a)-Mn-V's)  was based on the picture of Li$^{1+}$ charge state as in Zintl compounds,  Mn$^{2+}$ charge state as observed in (Ga,Mn)As and some other (III,Mn)V and (II,Mn)VI magnetic semiconductors, and the established close relationship between non-magnetic I(a)-II-V and III-V semiconductors.\cite{Jungwirth:2010_a} From this viewpoint, the Cu$^{1+}$Mn$^{2+}$Sb$^{3-}$ configuration inferred from previous density functional theory studies\cite{Jeong:2005_a} of this semi-Heusler alloy hints that the replacement of the I(a) alkali-metal with the I(b) Cu in I-Mn-V compounds is a promising approach. Although not a semiconductor, CuMnSb is also not a conventional metal with a high density of states at the Fermi energy but rather a semimetal with a small overlap between the bottom of the conduction band and the top of the valence band.\cite{Jeong:2005_a,Galanakis:2008_a} The relatively low room-temperature conductivity of $\sim 6\times10^3$~$\Omega^{-1}$cm$^{-1}$ measured in this compound supports the theoretically predicted semimetal band-structure.\cite{Boeuf:2006_a} By introducing a lighter group-V element As or P, one may expect the band-gap to fully open. 

%Based on full-potential density functional calculations of the metastable cubic (semi-Heusler) and stable orthorhombic phases we will discuss both the effects of chemical composition and crystal structure on the electronic properties of CuMnAs and CuMnP.

In the literature, CuMnSb is often quoted as a rare example among semi-Heusler alloys with AFM order.\cite{Forster:1968_a,Endo:1970_a,Abdulnoor:1980_a,Jeong:2005_a,Boeuf:2006_a,Duong:2007_a,Galanakis:2008_a} It has a N\'eel temperature of 50~K. Synthesis and crystal structure measurements of CuMnAs and CuMnP were reported in Ref.~\onlinecite{Mundelein:1992_a}. Unlike the cubic CuMnSb, the equilibrium crystal structure of CuMnAs and CuMnP is orthorhombic Pnma. Susceptibility measurements of CuMnAs presented in Ref.~\onlinecite{Mundelein:1992_a} indicate possible AFM order at room temperature.
%Again by considering both the cubic and orthorhombic phases of CuMnAs and CuMnP we inNaively, a tighter lattice arrangement resulting in stronger magnetic coupling can be expected when introducing the lighter group-V elements. The semimetal to semiconductor transition in the sequence CuMnSb$\rightarrow$CuMnAs$\rightarrow$CuMnP may be, therefore, also expected to be accompanied by an increasing  N\'eel temperature. 
To the best of our knowledge, the electronic and magnetic structure of CuMnAs and CuMnP has not been further investigated experimentally and no band structure calculations have been published to date.  In this paper we present full-potential density functional theory calculations of the electronic and magnetic structure of CuMnAs and CuMnP. We complement the theory analysis   by X-ray diffraction, magnetization, transport, and neutron diffraction measurements of chemically synthesized CuMnAs. As a reference, we also report our experimental results on CuMnSb.

\section{Theory}
To calculate the electronic structure of CuMn-V compounds we employed the full-potential linearized-augmented-plane-wave method (WIEN2k package).\cite{Wien2k}  Using the generalized-gradient approximation (GGA)\cite{Perdew:1996_a} we first compare energies of the FM and AFM states of the equilibrium orthorhombic crystal structure and the metastable cubic semi-Heusler structure of {CuMnAs} and {CuMnP}. In the former case we use experimental lattice parameters\cite{Mundelein:1992_a} $a = 6.5859$~\AA, $b = 3.8671$~\AA, and $c = 7.3202$~\AA\, for CuMnAs and $a = 6.3188$~\AA, $b = 3.7239$~\AA, and $c = 7.0883$~\AA\, for CuMnP. By minimizing the total energy of the cubic phase we determined the corresponding theoretical lattice constant $a=5.83$~\AA\, for CuMnAs and $a=5.66$~\AA\, for CuMnP. The calculated difference between the FM and AFM total energy in the cubic crystal phase is $E_{FM}-E_{AFM}=52$~meV/Mn-atom for CuMnAs and $E_{FM}-E_{AFM}=59$~meV/Mn-atom for CuMnP. Here we considered the same AFM alignment of Mn moments  in CuMnAs and CuMnP as reported in the neutron diffraction study of CuMnSb.\cite{Forster:1968_a} A comparison with theoretical results for CuMnSb, where $E_{FM}-E_{AFM}=50$~meV/Mn-atom,\cite{Jeong:2005_a} suggests  that $T_N$ increases when replacing Sb with As or P due to a tighter lattice arrangement. However, the expected increase of $T_N$ is only in the range of $\sim$20\% for the cubic phase. 

\begin{figure}[h!]
\hspace*{0cm}\epsfig{width=1\columnwidth,angle=0,file=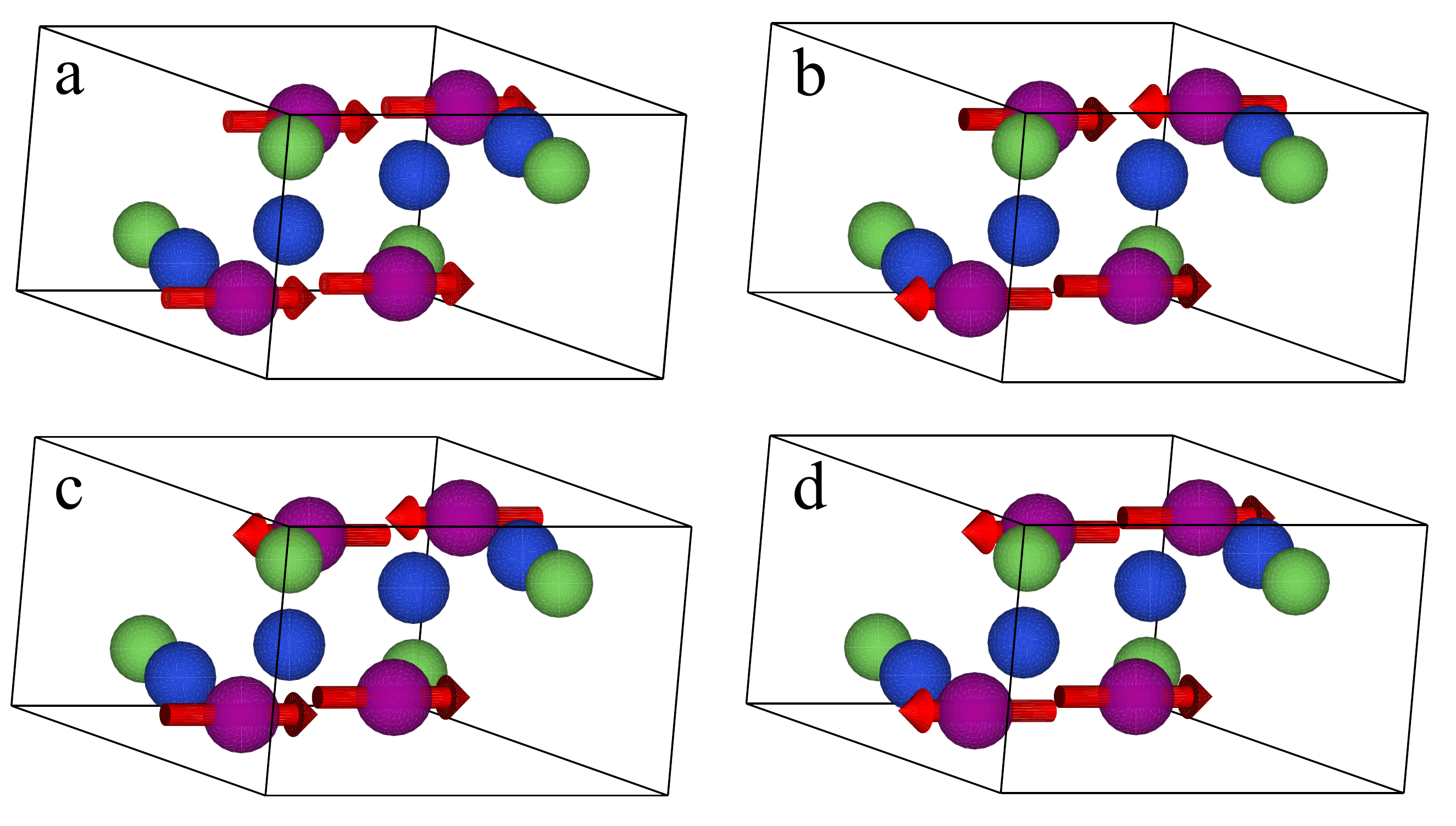}
\vspace*{-0.5cm}
\caption{Color on-line. Atomic and magnetic arrangements of the orthorhombic CuMnAs and CuMnP. Purple (large) spheres with arrows represent Mn, blue (medium) spheres Cu, and green (small) spheres As. (a) FM in-plane arrangement and FM coupling between the planes, (b) AFM in-plane arrangement and FM coupling between the planes, (c) FM in-plane arrangement and AFM coupling between the planes, (d) AFM in-plane arrangement and AFM coupling between the planes.
}
\label{cryststruct}
\end{figure}

For the orthorhombic structure we considered the FM and three different AFM moment configurations, as shown in Fig.~1. Mn atoms in these orthorhombic crystals are arranged in layers parallel to the $a-c$ plane and the different AFM arrangements can then be characterized by AFM in-plane arrangement and FM coupling between the planes (AFM$_{\rm IP}$-FM$_{\rm OP}$, Pn$^\prime$m$^\prime$a, Fig~1(b)), FM in-plane arrangement and AFM coupling between the planes (FM$_{\rm IP}$-AFM$_{\rm OP}$, Pn$^\prime$m$^\prime$a$^\prime$, Fig~1(c)), or AFM in-plane arrangement and AFM coupling between the planes (AFM$_{\rm IP}$-AFM$_{\rm OP}$, Pnm$^\prime$a, Fig~1(d)). In CuMnAs,   the GGA total energies of the FM$_{\rm IP}$-AFM$_{\rm OP}$ and AFM$_{\rm IP}$-AFM$_{\rm OP}$ states are very similar (the difference is 1~meV/Mn-atom) while the energy of the AFM$_{\rm IP}$-FM$_{\rm OP}$ state is significantly higher (by  80~meV/Mn-atom). In CuMnP, the AFM$_{\rm IP}$-FM$_{\rm OP}$ state has again the highest energy and the difference between the GGA energy of the states FM$_{\rm IP}$-AFM$_{\rm OP}$ and AFM$_{\rm IP}$-AFM$_{\rm OP}$ is 30~meV/Mn-atom with the latter AFM state having the lowest energy. When comparing  total energies of the FM and AFM states we took the  AFM$_{\rm IP}$-AFM$_{\rm OP}$ state, i.e. compared energies of the spin configurations in Fig.~1(a) and 1(d), and obtained $E_{FM}-E_{AFM}=241$~meV/Mn-atom for CuMnAs and $E_{FM}-E_{AFM}=250$~meV/Mn-atom for CuMnP. These values are significantly larger than in the cubic crystals and the theory, therefore, predicts  that the orthorhombic CuMnAs and CuMnP have significantly higher N\'eel temperature than the semi-Heusler CuMnSb. 
\begin{figure}[h!]
\hspace*{0cm}\epsfig{width=1\columnwidth,angle=0,file=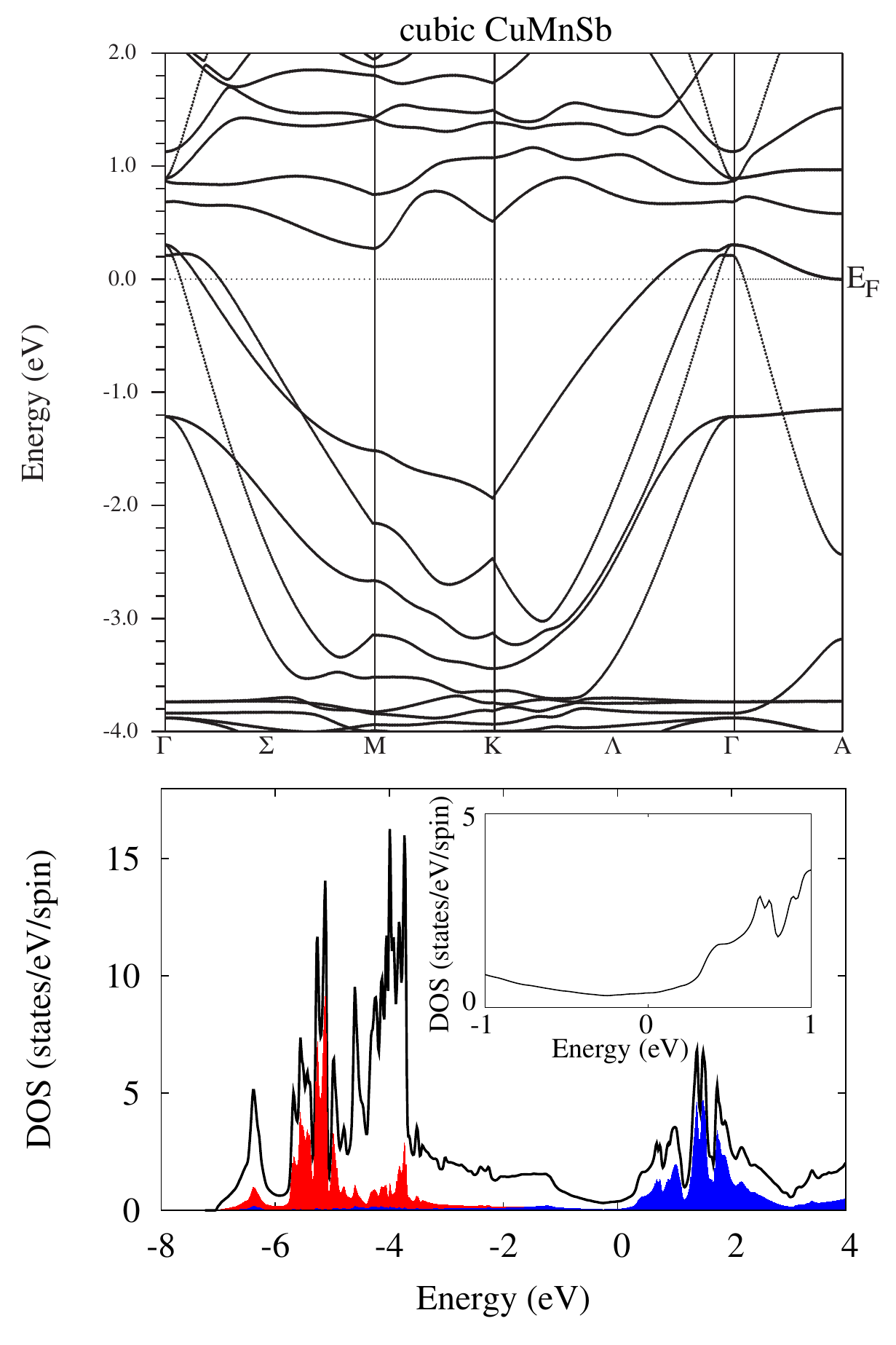}
\vspace*{-0.5cm}
\caption{Color on-line. GGA+U band structure plotted along lines in a hexagonal Brillouin zone (upper panel) and density of states (DOS) of cubic CuMnSb. The red  (filled lower energy) data correspond to the DOS projected onto  spin-up (spin-down) states of the 1st (2nd) Mn sublattice; the blue (filled higher energy) data correspond to the DOS projected onto spin-down (spin-up) states of the 1st (2nd) Mn sublattice.
}
\label{CuMnSb_cubic}
\end{figure}

The stability of the orthorhombic crystal phase of CuMnAs and CuMnP and the possibility for these compounds to form thin epitaxial films with the metastable cubic phase can be estimated from the difference in total energies of the two crystal structures. For CuMnAs we obtained $E_{cubic}-E_{ortho}=0.56$~eV/Mn-atom and for CuMnP the difference is $E_{cubic}-E_{ortho}=1.18$~eV/Mn-atom. For comparison, the total energy of the cubic phase of, e.g., GaN is larger than the energy of the equilibrium hexagonal phase by 0.02~eV/atom and in this case both crystal structures can be realized in thin films. For MnAs on the other hand, the difference between the cubic and hexagonal phases is 1~eV/Mn-atom and the cubic phase has been stabilized only in the form of nanocrystal inclusions in a cubic matrix.  From these comparisons we conclude that the orthorhombic phases of CuMnAs and CuMnP are very stable and the occurrence of the metastable cubic phases in bulk or thin film materials is unlikely.  

\begin{figure}[h!]
\hspace*{0cm}\epsfig{width=1\columnwidth,angle=0,file=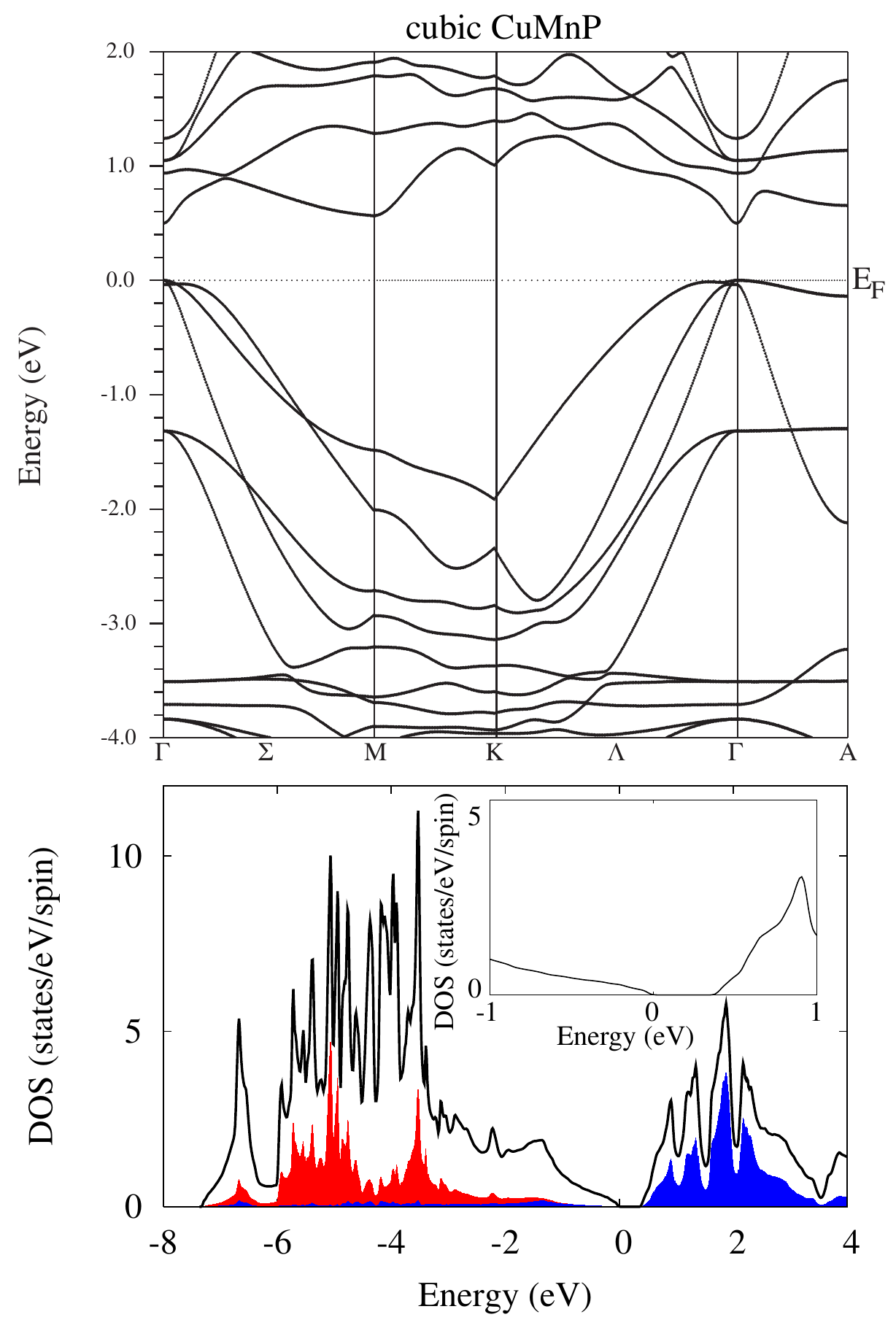}
\vspace*{-0.5cm}
\caption{Same as Fig.~\protect\ref{CuMnSb_cubic} for cubic CuMnP.}
\label{CuMnP_cubic}
\end{figure}
\begin{figure}[h!]
\hspace*{0cm}\epsfig{width=1\columnwidth,angle=0,file=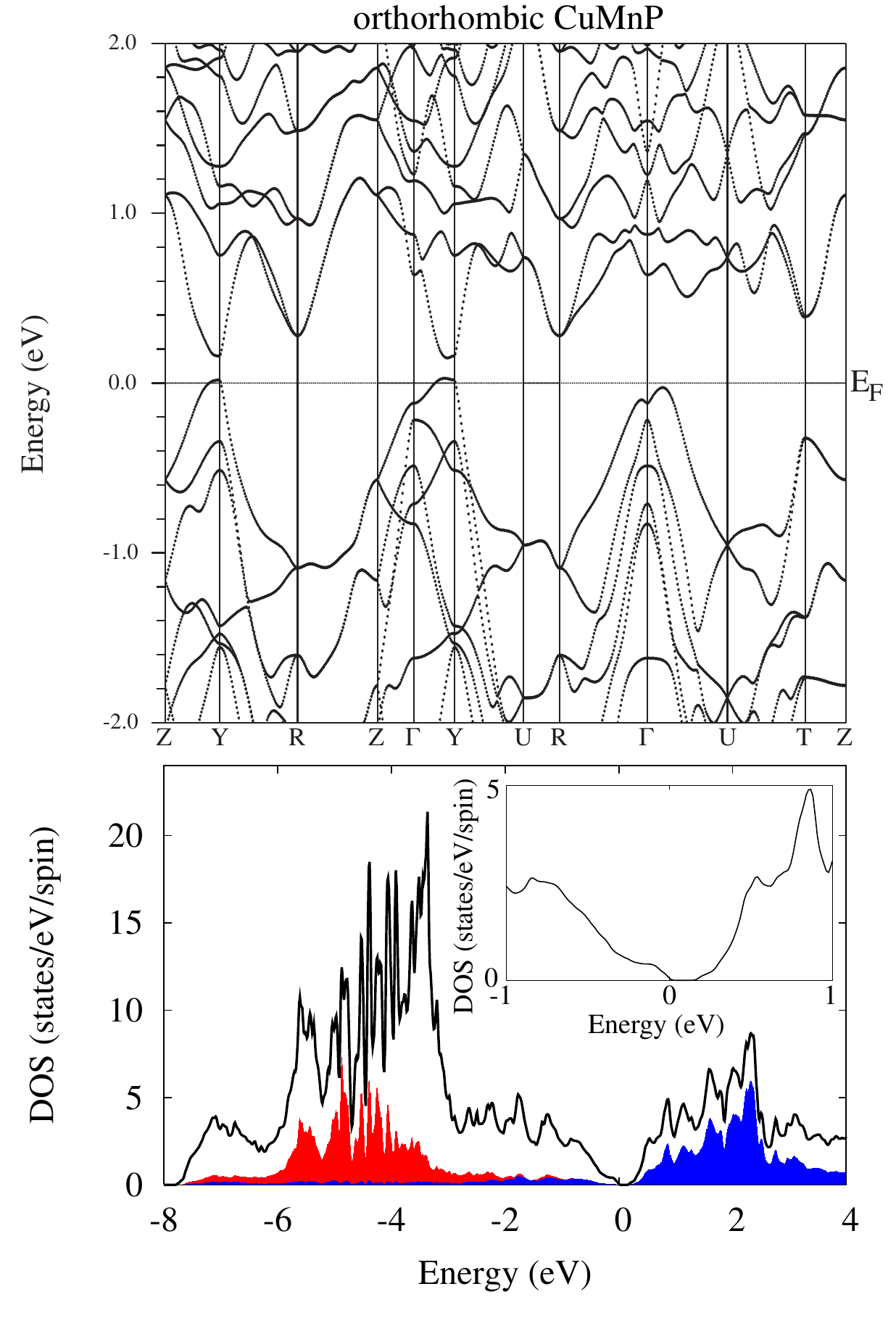}
\vspace*{-0.5cm}
\caption{Color on-line. GGA+U band structure (upper panel) and density of states (DOS) of orthorhombic CuMnP. The red  (filled lower energy) data correspond to the DOS projected onto  spin-up (spin-down) states of the 1st (2nd) Mn sublattice; the blue (filled higher energy) data correspond to the DOS projected onto spin-down (spin-up) states of the 1st (2nd) Mn sublattice.}
\label{CuMnP_ortho}
\end{figure}
\begin{figure}[h!]
\hspace*{0cm}\epsfig{width=1\columnwidth,angle=0,file=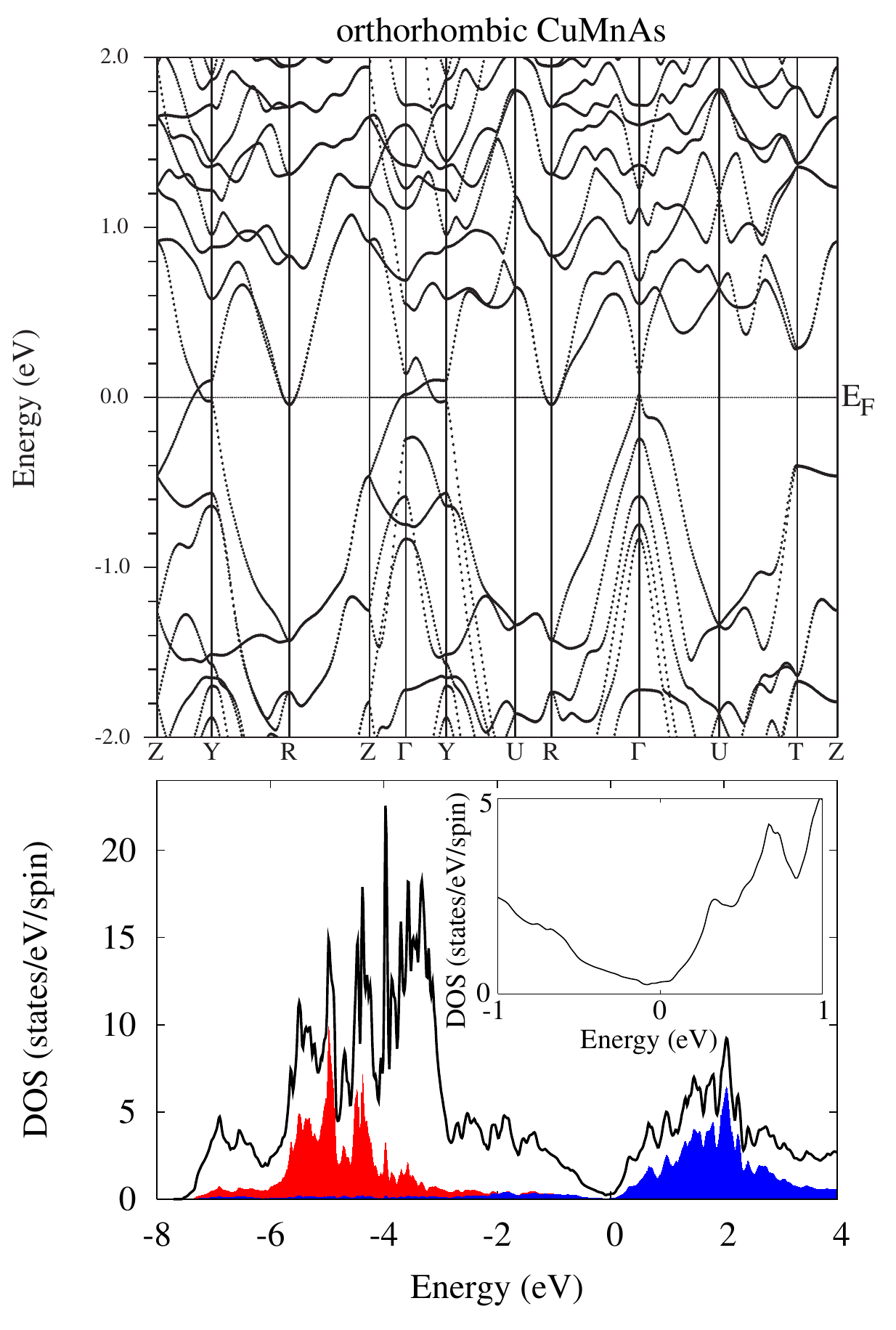}
\vspace*{-0.5cm}
\caption{Same as Fig.~\protect\ref{CuMnP_ortho} for orthorhombic CuMnAs.}
\label{CuMnAs_ortho}
\end{figure}
\begin{figure}[h!]
\hspace*{0cm}\epsfig{width=1\columnwidth,angle=0,file=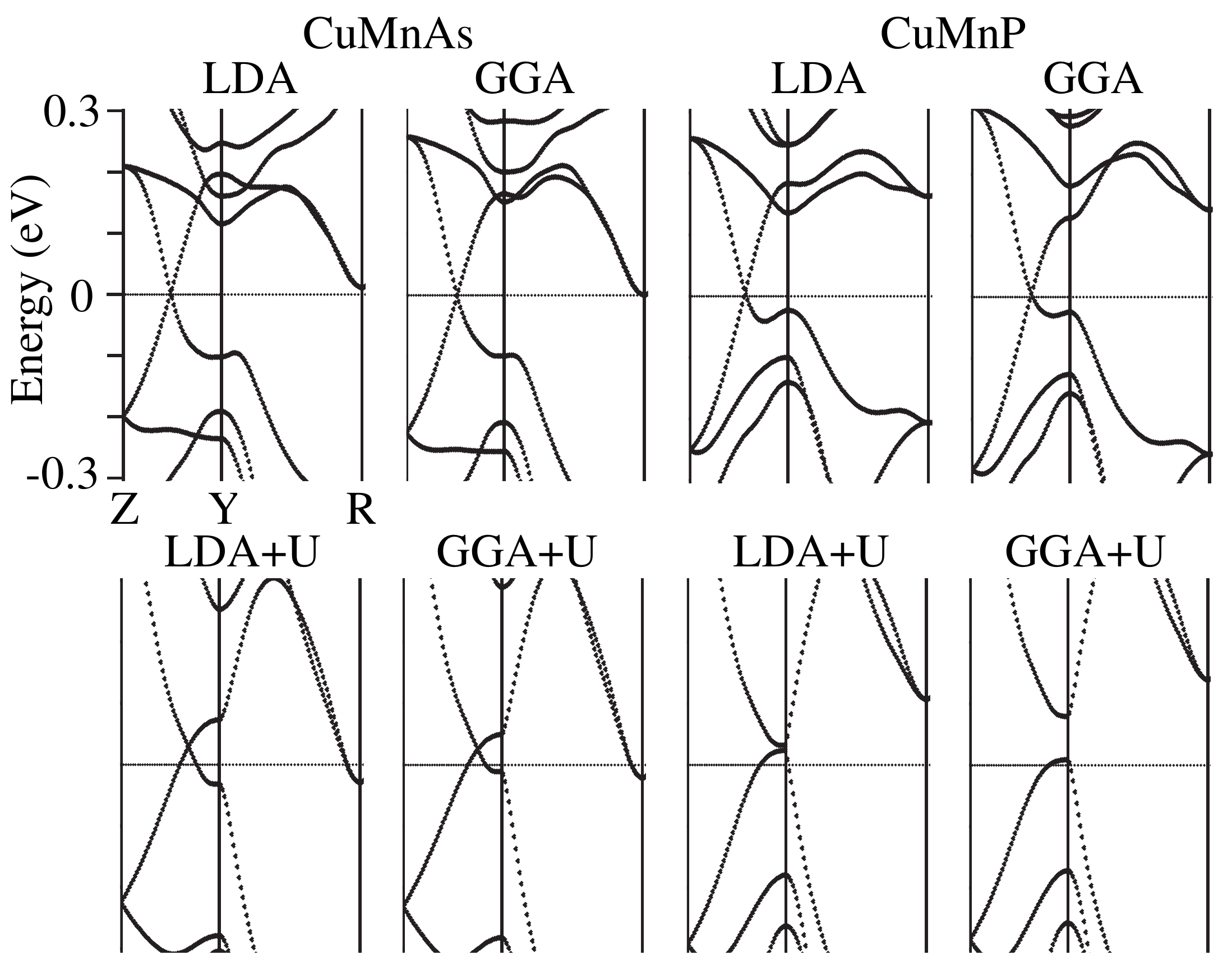}
\vspace*{-0.5cm}
\caption{Band structures of orthorhombic CuMnAs and CuMnP calculated using the depicted density-functional theory models.
}
\label{LDAGGA}
\end{figure}

Examples of the calculated band dispersions are shown in Figs.~\ref{CuMnSb_cubic}-\ref{LDAGGA}. In Figs.~\ref{CuMnSb_cubic} and \ref{CuMnP_cubic} we compare band structures of the cubic CuMnSb and CuMnP calculated in the GGA+U approximations with typical values of the correlation parameters\cite{Anisimov:1993_a,Liechtenstein:1995_a} for Mn $d$-orbitals,  $U=3.5$~eV and $J=0.6$~eV.  As already pointed out in Ref.~\onlinecite{Jeong:2005_a}, CuMnSb is a semimetal with a negative indirect band-gap. Fig.~\ref{CuMnP_cubic} shows that CuMnP, on the other hand, is already a semiconductor with a fully developed positive band-gap throughout the entire Brillouin zone. The semiconducting band structure of CuMnP is obtained also in the GGA+U calculation of the orthorhombic phase as shown in Fig.~\ref{CuMnP_ortho}. 

CuMnAs has a small but non-zero density of states at the Fermi energy in the GGA+U spectra as shown in Fig.~\ref{CuMnAs_ortho}. Since density functional theory tends to underestimate band gaps in semiconductors we conclude that the electronic structure of CuMnAs is in the transition region between a semimetal and a semiconductor. For completeness we compare in Fig.~\ref{LDAGGA} band structures of orthorhombic CuMnAs and CuMnP  calculated in the local density approximation (LDA),\cite{Perdew:1992_a} LDA+U, GGA, and GGA+U. In the plots we focus on the part of the spectra in which the gap opens in CuMnP. The spectra are consistent with common trends in III-V semiconductors of larger band gaps for lighter group-V elements and with the expected enhancement of the calculated semiconductor gap due to both the GGA correction to the local density theory and the correlation effects on Mn $d$-orbitals. Since the effects of these corrections are relatively strong, it might be desirable to employ in future studies computational methods which go beyond the GGA+U approximation in order to get a more quantitative understanding of the electronic structure of CuMn-V compounds. 

\section{Experiment}

In this section we discuss experimental properties of CuMnAs. Samples were prepared by direct synthesis from elements mixed in the stoichiometric 1:1:1 ratio using Cu (purity 99.999 \%), Mn (purity 99.98 \%), and As (purity 99.999 \%). They were placed into Al$_2$O$_3$ crucible and double sealed inside quartz ampoules. Samples were heated up to 1000$^\circ$C at a rate of 1$^\circ$C/min and annealed for 1 day. The reaction produced silver-gray solid rocks. Powder samples and polished-layer samples were prepared from  the material for X-ray diffraction measurements, chemical analysis, magnetization measurements by the superconducting quantum interference device (SQUID), van der Pauw transport measurements, and neutron diffraction measurements.  For comparison we also prepared CuMnSb samples using the same procedures.

\begin{figure}[h!]
\hspace*{0cm}\epsfig{width=1.1\columnwidth,angle=0,file=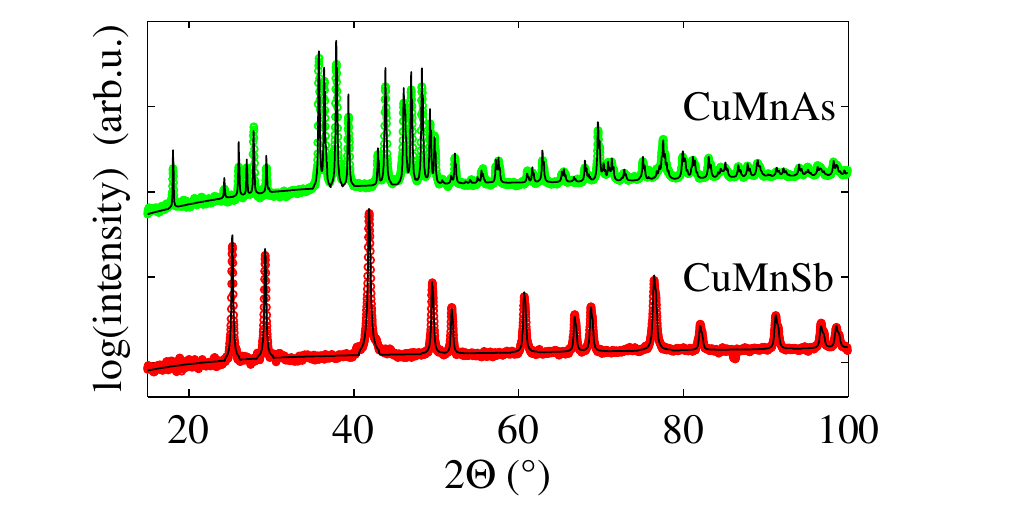}
\vspace*{-0.5cm}
\caption{X-ray diffraction data for CuMnAs and CuMnSb at room temperature. Green points are measured data of CuMnAs, red points are measured data of CuMnSb, and black thin lines are fits considering the orthorhombic crystal structure of CuMnAs and cubic crystal structure of CuMnSb.
}
\label{x-ray}
\end{figure}

In Fig.~\ref{x-ray} we show room-temperature X-ray measurements of our CuMnAs and CuMnSb materials. Diffraction patterns were obtained using a Bruker D8 Advance diffractometer in the Bragg-Brentano geometry and copper K$_\alpha$ radiation. Data analysis was carried out using the FullProf package.\cite{Carvajal:1993_a} The measurements show that CuMnAs has the orthorhombic Pnma crystal structure consistent with X-ray data in Ref.~\onlinecite{Mundelein:1992_a} and CuMnSb has the cubic semi-Heusler structure as reported previously.\cite{Endo:1970_a} Note that X-ray measurements of CuMnAs performed in the 300~K - 600~K range did not reveal any structural phase transition. Elemental analysis by energy-dispersive X-ray spectroscopy (EDX) confirmed the stoichiometric 1:1:1 ratio of Cu, Mn, and As(Sb) in the studied materials within the experimental error of $\sim$1\%.

\begin{figure}[h!]
\hspace*{-.8cm}\epsfig{width=1\columnwidth,angle=0,file=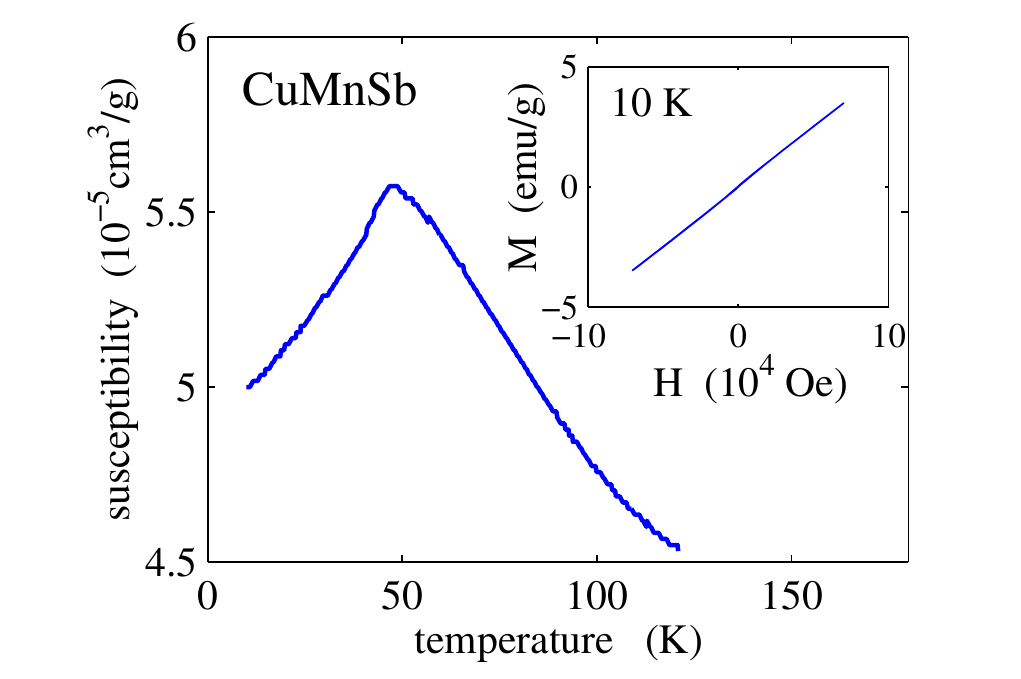}
\vspace*{-0.5cm}
\caption{Magnetic susceptibility at 5~kOe of CuMnSb. Inset shows magnetization measured at 10~K. 
}
\label{SQUIDCuMnSb}
\end{figure}

\begin{figure}[h!]
\hspace*{0cm}\epsfig{width=1\columnwidth,angle=0,file=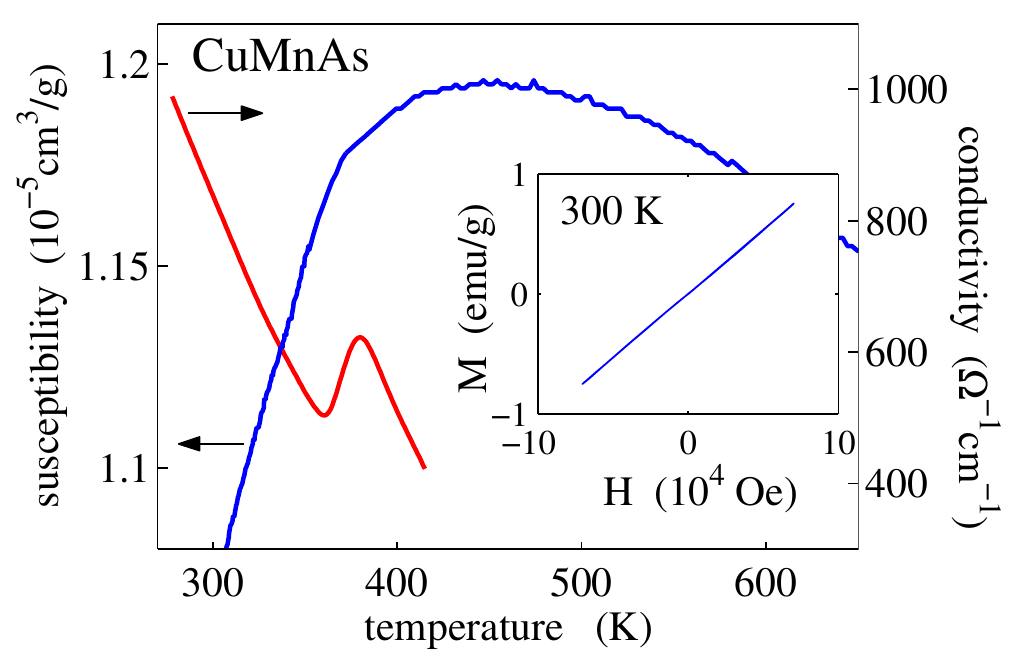}
\vspace*{-0.5cm}
\caption{Magnetic susceptibility at 5~kOe and conductivity of CuMnAs. Inset shows magnetization measured at 300~K. 
}
\label{SQUIDCuMnAs}
\end{figure}

In Fig.~\ref{SQUIDCuMnSb} we show measurements of the temperature dependent susceptibility and field dependent magnetization in CuMnSb. The data indicate $T_N\approx50$~K, in agreement with previous reports on this material.\cite{Endo:1970_a} Analogous measurements in CuMnAs are presented in Fig.~\ref{SQUIDCuMnAs}. We observe that the magnetic susceptibility of CuMnAs increases with increasing temperature up to much higher temperatures than in CuMnSb, in agreement with measurements presented in Ref.~\onlinecite{Mundelein:1992_a}. The dependence reverses above $\sim$450~K, indicating that $T_N$ is above room temperature.  As in the AFM CuMnSb we observe zero remanence below $T_N$ in CuMnAs (see inset in Fig.~\ref{SQUIDCuMnAs}). 

The transition to a magnetically ordered state in CuMnAs is observed also  in transport data shown in Fig.~\ref{SQUIDCuMnAs}. Above 400~K the material has a metallic conductance which increases with decreasing temperature. Near 400~K the conductance drops abruptly before it resumes the metallic character below 350~K. Such a conductance drop is commonly observed in magnetic materials and associated with critical scattering near the transition temperature. We also point out that the conductivity  at 300~K of CuMnAs is 800~$\Omega^{-1}$cm$^{-1}$ which is almost an order of magnitude lower than the room-temperature conductivity of the semimetal CuMnSb. This is consistent with the theoretically predicted trend towards semiconducting character of CuMn-V compounds with lighter group-V elements.

\begin{figure}[h!]
\hspace*{0cm}\epsfig{width=1\columnwidth,angle=0,file=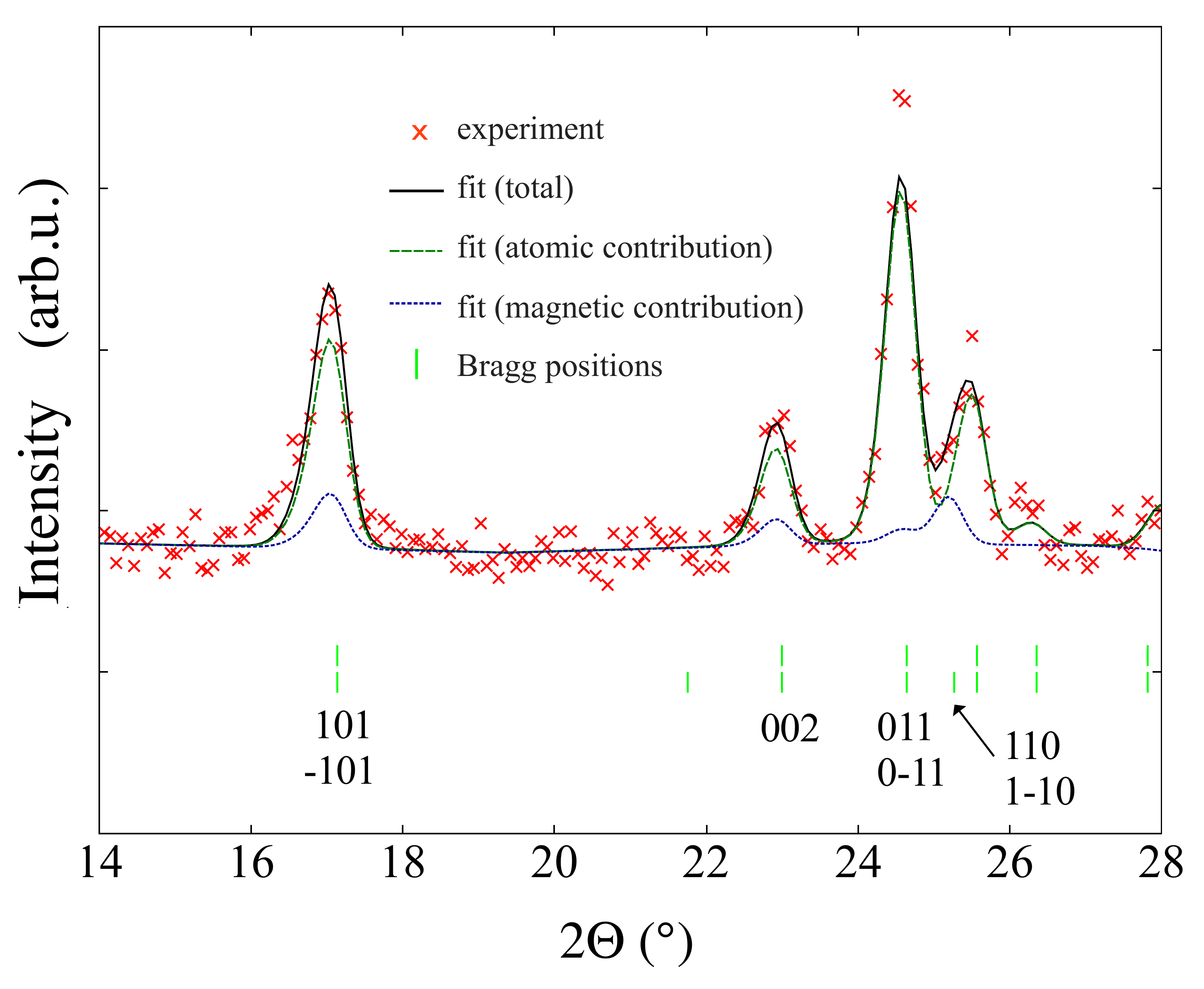}
\vspace*{-0.5cm}
\caption{Low angle part of measured and fitted neutron diffraction data of CuMnAs at room temperature. Atomic and magnetic contributions are plotted separately. Bragg positions of atomic (top) and magnetic (bottom) structures are also highlighted. Crystallographic indices denote the magnetic phase related  reflections. The peak in the magnetic contribution highlighted by arrow is characteristic of the AFM$_{\rm IP}$-AFM$_{\rm OP}$ Mn moment configuration.
}
\label{neutron}
\end{figure}

We conclude the experimental section by presenting initial results of room-temperature neutron diffraction measurements. The data shown in Fig.~\ref{neutron} were taken on the instrument MEREDIT in the Nuclear Physics Institute in Rez, Czech Republic. The neutron beam was monochromatized by a copper mosaic monochromator to a wavelength of 1.46 \AA. The data show  evidence of magnetic order at room temperature. The best fit using FullProf package\cite{Carvajal:1993_a} to the data was obtained for the  AFM$_{\rm IP}$-AFM$_{\rm OP}$ Mn moment configuration (Fig~1(d)).

\section{Summary}
In summary, we have performed a combined theoretical and experimental study of electronic and magnetic properties of CuMn-V compounds which complements  previous work on alkali-metal based I(a)-Mn-V AFM semiconductors. The transition from a low N\'eel temperature AFM CuMnSb to room-temperature AFMs CuMnAs and CuMnP is ascribed to the orthorhombic crystal structure of the latter two compounds. The transition from a semimetal CuMnSb to a semiconductor CuMnP is reminiscent of the trend of increasing band gap in III-V semiconductor compounds with lighter group-V elements. Since CuMn-V compounds are stable and readily compatible with common epitaxial growth techniques of high-quality semiconductor structures, they might represent favorable systems for exploring the concept of AFM-based spintronics.

\section*{Acknowledgment}
We thank  C. Frontera and V. Hol\'y for useful discussions and M. Mary\v{s}ko for experimental assistance and we acknowledge support  from EU Grant  FP7-215368 SemiSpinNet, and ERC Advanced Grant 268066, from Czech Republic Grants P204/11/P339, AV0Z10100520, AV0Z10100521, IAA100100912, LC510, and Preamium Academiae. 

%\bibliography{MSWEBpublications,other}

\end{document}